\documentclass{article}
\usepackage{spconf,amsmath,graphicx}
\usepackage{booktabs}

\title{Low-resource cross-domain singing voice synthesis via reduced self-supervised speech representations}
\name{
	\begin{tabular}{c}
		Panos Kakoulidis$^{\star}$,
		Nikolaos Ellinas$^{\star}$,
		Georgios Vamvoukakis$^{\star}$,
		Myrsini Christidou$^{\star}$,
		Alexandra Vioni$^{\star}$, \\ 
		Georgia Maniati$^{\star}$,
		Junkwang Oh$^{\dagger}$,
		Gunu Jho$^{\dagger}$,
		Inchul Hwang$^{\dagger}$,
		Pirros Tsiakoulis$^{\star}$,
		Aimilios Chalamandaris$^{\star}$
	\end{tabular}
}
\address{$^{\star}$ Innoetics, Samsung Electronics, Greece \\
	$^{\dagger}$ Mobile Communications Business, Samsung Electronics, Republic of Korea}

\begin{document}
%
\maketitle
\begin{abstract}
In this paper, we propose a singing voice synthesis model, Karaoker-SSL, 
that is trained only on text and speech data as a typical multi-speaker acoustic model. 
It is a low-resource pipeline that does not utilize any singing data end-to-end, 
since its vocoder is also trained on speech data. Karaoker-SSL is conditioned by 
self-supervised speech representations in an unsupervised manner. We preprocess
these representations by selecting only a subset of their task-correlated dimensions. 
The conditioning module is indirectly guided to capture style information during 
training by multi-tasking. This is achieved with a Conformer-based module, which 
predicts the pitch from the acoustic model's output. Thus, Karaoker-SSL allows 
singing voice synthesis without reliance on hand-crafted and domain-specific 
features. There are also no requirements for text alignments or lyrics timestamps. 
To refine the voice quality, we employ a U-Net discriminator that is conditioned 
on the target speaker and follows a Diffusion GAN training scheme.

\end{abstract}
\section{Introduction}
\label{sec:intro}

Despite the recent advancements in speech synthesis with deep learning,
modeling a melodic human voice remains an open challenge \cite{challenge}. 
Singing voice synthesis (SVS) is the task of generating vocals for a target 
speaker that adhere to a template audio or a music score. Predicting how a 
human sounds beyond the standard operation of the vocal tract is a complex 
objective. Singing is a controlled and structured voice production, which is
characterized by high pitch variation and other vocal manipulations, altering  
the speaker's identity. For this reason, SVS is an edge and complex case of 
human voice synthesis and it follows the progress of normal speech synthesis
research.

\subsection{Related Work}
The current state of the art in text-to-speech (TTS) involves transformer-based 
neural networks with flow-based decoders for acoustic modeling \cite{vits}. Some 
variations either incorporate modules that separately model vocal characteristics  
such as pitch and energy, to enrich the synthesized voice \cite{fastspeech2} or 
utilize self-supervised learning (SSL) speech representations \cite{saltts}. 
Explicit duration modeling has been established, replacing attention-based 
alignments. Expressive TTS and voice conversion (VC) extend these approaches 
\cite{periodvits, acevc}. Recently, a new generation of speech synthesis models 
have emerged, which leverage neural codecs and the diffusion-based training scheme 
\cite{naturalspeech2,vallex}.

The latest efforts in singing voice synthesis (SVS) build upon this progress, 
introducing adapted models for this task that process additional music information. 
These solutions are either two-way \cite{xiaoice, rmssinger, diffsinger} or end-to-end 
\cite{naturalspeech2, fastsvc, visinger2}. They rely on domain-specific data such as 
singing recordings, music scores or/and lyric alignments. However, these resources 
require time-consuming and expensive labor with error-prone results, complicating 
the deployment and scaling of dependent systems in a production setting.

\subsection{Proposed method}

In our previous work, we developed Karaoker \cite{karaoker}, a conditioned 
sequence-to-sequence acoustic model. It is designed according to a constrained task 
of SVS, where only unaligned text and speech audios are available. Also, its design 
aims to produce a lightweight but robust final subsystem by incorporating multiple 
modules that are active only in training. The vocoder that is used in the pipeline 
is similarly trained only on speech data.

We present Karaoker-SSL, which is based on the previous Karaoker principles, incorporating 
some of the recent advancements discussed above. The contributions of this work are listed below:

\begin{itemize}
	    \item We utilize a parallel speech-singing dataset to select the most relevant 
	    dimensions of SSL speech representations for the SVS task.
		\item We condition an acoustic model in an unsupervised manner with the reduced SSL 
		speech representations.
		\item The acoustic model predicts the pitch from its decoder outputs, performing 
		multi-tasking.
		\item The predicted pitch is evaluated with a custom loss of multiple comparisons.
		\item The generated mel spectrogram is evaluated by a U-Net discriminator
		using differentiable augmentations.
\end{itemize}

\section{METHOD}
\label{sec:method}

\subsection{SSL Dimension selection}
\label{ssec:dimselect}

\begin{figure}[t]
	\centering
	\includegraphics[width=70mm]{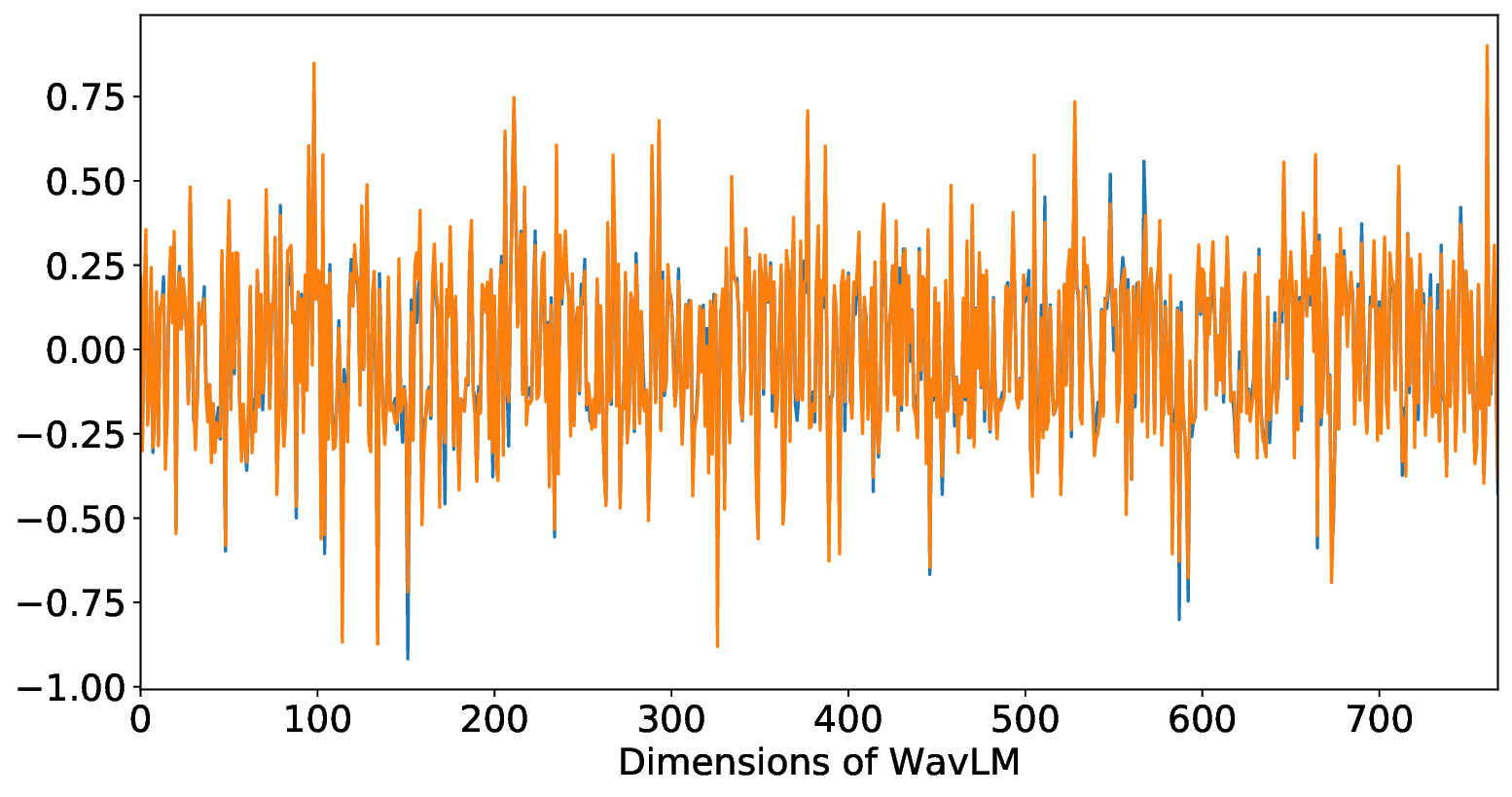}
	\caption{Overlapped WavLM embeddings (768 dim.), each one averaged on the time dimension, for two parallel audios of the same speaker and content. Blue is for normal speech and orange is for singing.}
	\label{fig:meanemb}
\end{figure}

Managing large representations is a challenge in terms of resources for the training and 
deployment of a model. We propose a task-specific strategy that effectively reduces the 
dimensions of a speech-based SSL representation by $\sim$88\%. This reduction also aims 
to keep the style information and minimize the linguistic content. At first, we extract 
SSL embeddings (Figure \ref{fig:meanemb}) from an internal multi-speaker parallel 
dataset, which includes 985 double takes (speech/singing) of one well-known verse. Then, 
we compute the absolute differences between the embeddings of the parallel audios of each 
speaker, averaged on the time dimension. We choose the dimensions that have median absolute 
differences across the dataset of z-score \(\geq 1\). This strategy is applied on  Wav2Vec 
2.0 \cite{wav2vec2} and WavLM \cite{wavlm} embeddings and we choose the latter as it leads 
to better quality.

\subsection{Acoustic modeling}
\label{ssec:arch}

\begin{figure}[t]
	\centering
	\includegraphics[width=60mm]{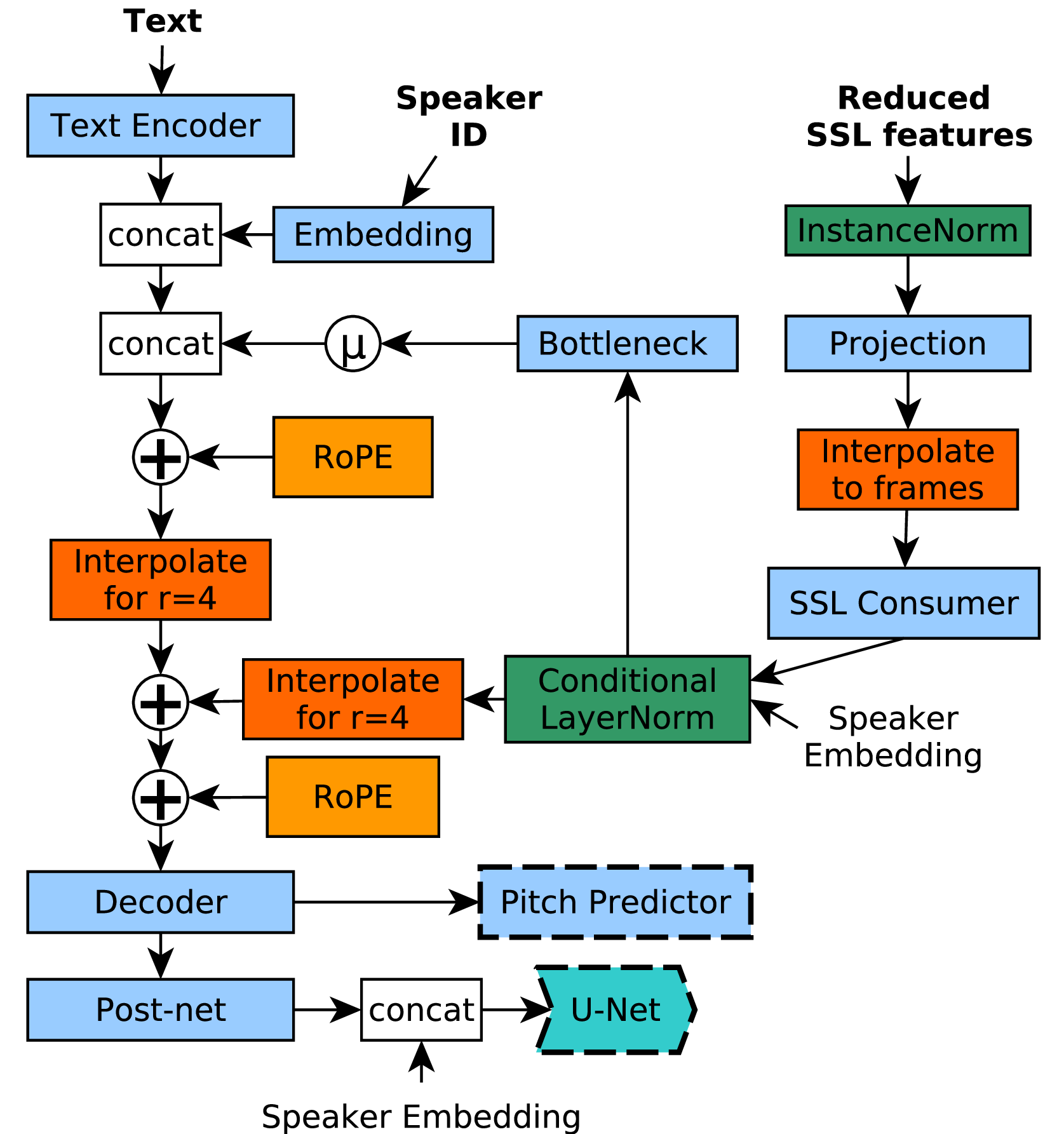}
	\caption{Architecture of Karaoker-SSL}
	\label{fig:architecture}
\end{figure}

The acoustic model is a non-attentive version of the one used in \cite{karaoker} (Figure 
\ref{fig:architecture}). The encoder outputs are firstly concatenated with speaker and global 
SSL embeddings. Then, they are summed with rotary positional embeddings (RoPE) \cite{rotary} 
and expanded to the maximum mel length by Nearest Neighbours interpolation. Local SSL 
embeddings, that undergo a similar process of length interpolation and RoPE  addition, 
are added to the expanded encoder outputs and then passed altogether to the decoder.
The SSL embeddings are produced by SSL Consumer that conditions the acoustic model.

\subsection{SSL Consumer}
\label{ssec:consume}

At first, the reduced WavLM embeddings, rWavLM, are projected to 8 dimensions with a linear 
layer. The projection is followed by instance normalization \cite{instancenorm}, Mish activation 
\cite{mish} and linear interpolation to mel length. Then, they are provided to SSL Consumer, 
which consists of 14 convolutional blocks as the ones used in the feature encoder of 
\cite{karaoker} but with Mish as the activation function. The first 6 blocks do not have 
an activation in their final layer and have a kernel size of 3. The outputs are normalized 
with Conditional Layer Normalization \cite{adaspeech} on speaker embeddings. An 1D convolutional 
layer (1x1) is used as a bottleneck on the SSL Consumer's outputs for the learning of global SSL 
embeddings. The output of this layer is averaged on the time dimension to facilitate the training
of SSL Consumer as an InfoVAE \cite{infovae}.  The global embeddings are also trained to exhibit 
minimum mutual information with the encoder outputs.

\subsection{Multi-tasking}
\label{ssec:multit}

Except for the main objective of mel spectrogram generation, Karaoker-SSL has an auxiliary
task of predicting the pitch from the decoder's outputs. A four-layered Conformer \cite{conformer}
was employed for these predictions with a 128-dimensional feedforward layer, 8 multi-attention 
heads and kernel size of 15 for its depthwise convolution layers. The mel spectrogram is projected 
to 64 dimensions with a linear layer before being sent to the Conformer. The pitch predictor's
outputs are projected to a single dimension and they are compared with the ground truth pitch. 

In our early experiments, we observe that known regression or alignment metrics, such as mean 
squared error (MSE) and differentiable dynamic time warping (DTW), are proved to be less effective 
for the task. We suggest the following loss function for pitch comparison:

\begin{multline}
	L_{pitch} = ln \bigg(\sum_{i=1}^{N}\bigg(\Big\Vert DFT(means_{w}(F0_{gt}, div_{i})) - \bigg. \bigg.\\
	\bigg. \bigg. DFT(means_{w}(F0_{gen}, div_{i}))\Big\Vert_{1}\bigg)\bigg)
\end{multline}

Where \( N \) are the number of mel length's proper divisors \(div_{i}\) (\(div_{i} > 1,  \forall 
\,\, 1 \leq i \leq N\)), \(DFT\) is the Discrete Fourier Transform, \(means_{w}\) is the series 
of the windowed means by a window size of \(div_{i}\), \(F0_{gt}\) is the ground truth pitch
and \(F0_{gen}\) is the predicted pitch. This function produces and compares multiple smoothed 
versions of the two pitch contours with minimal overhead.  

\subsection{U-Net Discriminator}
\label{ssec:unetd}

We implemented a U-Net \cite{unet} with convolutional blocks of two 1D convolutional 
layers (kernel size=3, padding=1) which are accompanied with spectral normalization 
\cite{spectralnorm} and GeLU \cite{gelu} activation. The downsampling path (64 to 512 
dimensions) consists of 4 blocks with 20\% dropout and max pooling (kernel size=2, 
stride=2). Its symmetric upsampling path (512 to 64 dimensions) includes 4 blocks, 
each one preceded by a transposed convolutional layer (kernel size=2, stride=2). 

The idea of using a U-Net as a discriminator has been mentioned recently for wave 
signals \cite{unetd}. Here, we employ U-Net to discriminate random slices of mel 
spectrograms \cite{karaoker} by the output of the last downsampling block. The discriminator is 
trained as in \cite{karaoker} but only with combined augmentations of SpecAugment 
\cite{specaugment}, Diffusion GAN \cite{diffusiongan} and real-as-fake \cite{realfake}.
Thus, the U-Net is not trained on clean ground truth data. Also, mel reconstruction 
($L_{1} + L_{2}$ losses) has been added in the dicriminator's training objective.

\subsection{Losses}
\label{ssec:loss}

The training objective of the Karaoker-SSL consists of the following losses:

\begin{equation}
	\begin{split}
		L = L_{mel} + \beta L_{mmd} +  \kappa L_{mi} + L_{pitch} \\ 
		    +  \lambda L_{pitch\_g} +  L_{pitch\_repr} - L_{D}
		\label{eq2} 
	\end{split}
\end{equation}

\noindent where $L_{mel}$ is the sum of $L_{1}$, $L_{2}$ losses between  ground truth and
decoder/post-net outputs, $L_{mmd}$ is Maximum Mean Discrepancy for InfoVAE's training, 
$L_{mi}$ is the Mutual Information minimization loss, $L_{pitch}$ and $L_{pitch\_g}$ refer
to pitch prediction from ground truth and generated mels, $L_{pitch\_repr}$ is the application
of the custom pitch loss function between the intermediate 64-dim. pitch learning representations 
of ground truth and generated mels and $L_{D}$ is the discriminator's loss that should be 
maximized.

\section{Experiments}
\label{sec:experiments}

\subsection{Experimental setup}
\label{ssec:setup}

We perform pitch extraction in the same manner as in \cite{karaoker} from the female 
voices in LibriTTS \cite{libritts} and VCTK \cite{vctk} (VCTK+Libri-f). The pitch is divided 
by 100 and random noise is applied ($\pm$ 0.01). In this setup, we process audio files with 
24KHz sampling rate as accepted by the efficient LPCNet \cite{lpcnet}, which is trained on 
an internal speech dataset.

The whole model has 12M parameters and it is trained for 600k iterations with batch size 
of 16 samples, $\beta$=0.1 and $\kappa$=10000. The pitch predictor infers the pitch of the generated mel 
spectrograms with $\lambda$=100 after 100k iterations of training. The discriminator starts 
learning at 150k iterations and discriminates at 250k iterations. Two separate Adam optimizers \cite{adam} 
are used for the generator (acoustic model) and the discriminator with a starting learning 
rate of 0.001. The rate of the first optimizer drops in half every 100k iterations, while 
the second optimizer follows a step learning scheduler (gamma=0.1). Concerning inference, 
only the unaligned text and a rWavLM representation are required, which form the template 
for the SVS.

\subsection{Subjective evaluation}
\label{ssec:subjective}

We conduct three types of listening tests in order to evaluate the proposed model. We survey
native speakers in  by Mean Opinion Score (MOS) (37 testers), Speaker Similarity (spk-sim) 
(41 testers) and Song Similarity (song-sim) (29 testers) as described in \cite{karaoker}. 
For reproducibility, we evaluate Karaoker-SSL on  the combined public dataset VCTK+Libri-f 
and we investigate the suitability of the reduced Wav2Vec 2.0 (rWav2Vec) and rWavLM embeddings for 
cross-domain SVS. The listening tests are applied on 10 source tracks of different length, 
singer's gender, quality and melodic complexity. 

We notice that the model performs better with rWavLM embeddings (Table \ref{table:femalemos}) 
and is robust enough to process noisy vocals from source separation predictions. Also, we observe that the
 model attempts to reproduce the vocal manipulations of the source, distorting speaker identity
(Tables \ref{table:femalemos},\ref{table:targetmos}). Except for potential technical shortcomings 
such as lack of data coverage, the speaker similarity can be affected from the fact that a person's 
singing style deviates from her normal speaking style. From the breakdown of the scores per speaker, 
we confirm that the model's outputs are consistent even for target speakers with different data 
coverage  (Table \ref{table:targetmos}). The high standard deviation values emphasize the difficulty 
of testing SVS as we mention in \cite{karaoker}.

\begin{table}[htb]
	\caption{The results of listening tests for the conditioning of   Karaoker-SSL by rWav2Vec
		(89 dim.) and rWavLM (93 dim.). The target speakers are: 
		s5, p276, p362, p265, p234 (VCTK+Libri-f dataset).}
	\label{table:femalemos}
	\centering
	\vspace{3mm}
	\begin{tabular}{llll}
		\toprule
		&  \multicolumn{3}{c}{\textbf{Scores}}  \\
		\textbf{Embedding} &  \textbf{MOS} &  \textbf{spk-sim}  &  \textbf{song-sim}\\
		\midrule

		rWav2Vec  & 3.32 $\pm$ 0.96 &  2.66 $\pm$ 0.87  & 2.79 $\pm$ 1.1\\
		
		rWavLM & \textbf{3.44 $\pm$ 0.94} & \textbf{2.81 $\pm$ 0.87} & \textbf{3.17$\pm$ 0.95} \\	
		\bottomrule
	\end{tabular}
\end{table}

\begin{table}[htb]
	\caption{The results of listening tests per target speaker with different training data durations 
	for Karaoker-SSL trained on the VCTK+Libri-f dataset.}
	\label{table:targetmos}
	\centering
	\vspace{3mm}
	\begin{tabular}{llll}
		\toprule
		&  \multicolumn{3}{c}{\textbf{Scores}}  \\
		\textbf{Target (data)} &  \textbf{MOS} &  \textbf{spk-sim}  &  \textbf{song-sim}\\
		\midrule
		p234 (19 min)  & 3.51 $\pm$ 0.95 & 2.86 $\pm$ 0.83  & 3.13 $\pm$ 0.96\\
		p265 (20 min) & 3.29 $\pm$ 1.02 & {2.82 $\pm$ 0.84} & {3.18 $\pm$ 0.92} \\	
		p276 (20 min) & {3.45 $\pm$ 0.97} & {2.75 $\pm$ 0.91} & {3.21 $\pm$ 0.90} \\	
		p362 (5 min) & {3.42 $\pm$ 0.91} & {2.85 $\pm$ 0.81} & {3.00 $\pm$ 0.96} \\	
		s5 (20 min ) & {3.54 $\pm$ 0.84} & {2.78 $\pm$ 0.97} & {3.31 $\pm$ 1.0} \\	
		\bottomrule
	\end{tabular}
\end{table}

\subsection{Objective evaluation}
\label{ssec:objective}

Our objective evaluation is conducted as in \cite{karaoker} using the template data from the listening tests.
It yields that the synthesized voice has a high median speaker embedding similarity of \textbf{70.2\%} and \textbf{0.26} \textbf{mF0 RMSE} (median-normalized pitch Root Mean Square Error). We also compute
\textbf{mF0 NRMSD} (Normalized Root Mean Square Deviation) which is \textbf{18.2\%}. Although the latter
value is far from perfect, it describes a fair accuracy in terms of the objective evaluation, given the 
difficulty of the task and since there is no other model available for direct comparison. Equivalent claims
have been made in our previous work \cite{karaoker}.

\subsection{Ablations}
\label{ssec:ablations}

We investigate how each auxiliary module contributes to the acoustic model, thus we prepare
some ablations. The baseline (ablation\{1\}) is only the conditioned acoustic 
model and the ablations are ablation\{2\} ('nop'), which lacks a pitch predictor, 
and the U-Net-missing ablation\{3\} ('nod'). They are jointly tested with the other 
systems in the listening test of \ref{ssec:subjective}. From the results in Table \ref{table:ablationmos}, 
it is evident that the pitch predictor and the discriminator are critical for the task 
(ablation \{4\}). 

Both components elevate the capabilities of the acoustic model for SVS. The discriminator 
contributes to song similarity as it might aid the auxilliary task of pitch prediction, 
apart from the main task of mel spectrogram generation. Essentially, the acoustic model is the 
only active network during inference, thus the final production model for each ablation and 
Karaoker-SSL is the same. However, Karaoker-SSL's acoustic model performs better 
due to its enriched representations as a result of its multi-objective training.

\begin{table}[htb]
	\caption{The results of listening tests on the ablations for target speakers s5, p276, p362, p265, 
	p234 in the VCTK+Libri-f dataset.}
	\label{table:ablationmos}
	\vspace{3mm}
	\centering
	\begin{tabular}{llll}
		\toprule
		&  \multicolumn{3}{c}{\textbf{Scores}}  \\
		\textbf{Ablation\{No\}} &  \textbf{MOS} &  \textbf{spk-sim} & \textbf{song-sim} \\
		\midrule
		base\{1\}   & 3.41 $\pm$ 0.98 & 2.79 $\pm$ 0.83  & 3.01 $\pm$ 1.09\\
		nop\{2\}  & 3.42 $\pm$ 0.92 &  2.79 $\pm$ 0.85  & 3.14 $\pm$ 1.01\\
		nod\{3\}   &\textbf{3.56 $\pm$ 0.89} & 2.75 $\pm$ 0.85  & 3.10 $\pm$ 0.98\\
	    full\{4\} & {3.44 $\pm$ 0.94} & \textbf{2.81 $\pm$ 0.87} & \textbf{3.17$\pm$ 0.95} \\	
		\bottomrule
	\end{tabular}
\end{table}

\subsection{Comparison with Karaoker}
\label{ssec:karaoker}

Although the scores in \cite{karaoker} are not directly comparable due to the different conditioning, 
template data, sampling rate, vocoder and target speakers, we could assume that Karaoker-SSL produces 
a more natural sounding result. The song similarities of the two models are comparable, even though 
Karaoker-SSL uses a fraction of the hand-crafted features during training and none during inference. 
We fail to detect any recent solution that approaches the SVS task by using only speech data end-to-end 
for its training. Therefore, we encourage the reader to listen to the demo samples of Karaoker-SSL\footnote{ https://innoetics.github.io/publications/karaoker-ssl/index.html \label{demopage}} and make his/her 
own judgement.

\section{Conclusion}
\label{sec:conclusion}

We have presented Karaoker-SSL, a model that can synthesize singing voice
without having seen any singing data in its own training.  We selected the
dimensions of SSL speech representations according to the SVS task by analyzing
a parallel speech-singing dataset. The model leverages these reduced reprentations
to condition acoustic modeling without direct supervision. Unsupervised conditioning
serves the scenario of an easily scalable and maintainable production model with
low demand on resources. We observe that the acoustic model's co-training with 
auxilliary modules greatly enhances the overall performance towards 
the SVS task, without any increase in the number of the production model's parameters. 
Also, a level of disentaglement is achieved between  linguistic and acoustic 
information. Yet, it is not enough for achieving full controllability on the generated 
voice by the input text and it will be prioritized in our future focus.

\label{sec:refs}

\bibliographystyle{IEEEbib}
\ninept
\bibliography{refs}

\begin{thebibliography}{10}

\bibitem{challenge}
W.-C. Huang, L.~P. Violeta, S. Liu, J. Shi, Y. Yasuda, and T. Toda,
\newblock ``{The Singing Voice Conversion Challenge 2023},''
\newblock {\em arXiv preprint arXiv:2306.14422}, 2023.

\bibitem{vits}
J. Kim, J. Kong, and J. Son,
\newblock ``{Conditional Variational Autoencoder with Adversarial Learning for
  End-to-End Text-to-Speech},''
\newblock in {\em Proc. ICML}, 2021.

\bibitem{fastspeech2}
Y. Ren, C. Hu, X. Tan, T. Qin, S. Zhao, Z. Zhao, and T.-Y. Liu,
\newblock ``{FastSpeech 2: Fast and High-Quality End-to-End Text to Speech},''
\newblock in {\em Proc. ICLR}, 2021.

\bibitem{saltts}
R. Sivaguru, V.~S. Lodagala, and S. Umesh,
\newblock ``{SALTTS: Leveraging Self-Supervised Speech Representations for
  improved Text-to-Speech Synthesis},''
\newblock in {\em Proc. Interspeech}, 2023.

\bibitem{periodvits}
Y. Shirahata, R. Yamamoto, E. Song, R. Terashima, J.-M. Kim, and K. Tachibana,
\newblock ``{Period VITS: Variational Inference with Explicit Pitch Modeling
  for End-To-End Emotional Speech Synthesis},''
\newblock in {\em Proc. ICASSP}, 2023.

\bibitem{acevc}
S. Hussain, P. Neekhara, J. Huang, J. Li, and B. Ginsburg,
\newblock ``{ACE-VC: Adaptive and Controllable Voice Conversion Using
  Explicitly Disentangled Self-Supervised Speech Representations},''
\newblock in {\em Proc. ICASSP}, 2023.

\bibitem{naturalspeech2}
K. Shen, Z. Ju, X. Tan, Y. Liu, Y. Leng, L. He, T. Qin, S. Zhao, and J. Bian,
\newblock ``{NaturalSpeech 2: Latent Diffusion Models are Natural and Zero-Shot
  Speech and Singing Synthesizers},''
\newblock {\em arXiv preprint arXiv:2304.09116}, 2023.

\bibitem{vallex}
Z. Zhang, L. Zhou, C. Wang, S. Chen, Y. Wu, S. Liu, Z. Chen, Y. Liu, H. Wang,
  J. Li, L. He, S. Zhao, and F. Wei,
\newblock ``{Speak Foreign Languages with Your Own Voice: Cross-Lingual Neural
  Codec Language Modeling},''
\newblock {\em arXiv preprint arXiv:2303.03926}, 2023.

\bibitem{xiaoice}
W. Chunhui, C. Zeng, and X. He,
\newblock ``{XiaoiceSing 2: A High-Fidelity Singing Voice Synthesizer Based on
  Generative Adversarial Network},''
\newblock in {\em Proc. Interspeech}, 2023.

\bibitem{rmssinger}
J. He, J. Liu, Z. Ye, R. Huang, C. Cui, H. Liu, and Z. Zhao,
\newblock ``{RMSSinger: Realistic-Music-Score based Singing Voice Synthesis},''
\newblock in {\em Findings of the Association for Computational Linguistics:
  ACL 2023}, 2023.

\bibitem{diffsinger}
J. Liu, C. Li, Y. Ren, F. Chen, and Z. Zhao,
\newblock ``{DiffSinger: Singing Voice Synthesis via Shallow Diffusion
  Mechanism},''
\newblock {\em Proc. AAAI}, vol. 36, no. 10, 2022.

\bibitem{fastsvc}
S. Liu, Y. Cao, N. Hu, D. Su, and H. Meng,
\newblock ``{FastSVC: Fast Cross-Domain Singing Voice Conversion With
  Feature-Wise Linear Modulation},''
\newblock in {\em Proc. ICME}, 2021.

\bibitem{visinger2}
Y. Zhang, H. Xue, H. Li, L. Xie, T. Guo, R. Zhang, and C. Gong,
\newblock ``{VISinger2: High-Fidelity End-to-End Singing Voice Synthesis
  Enhanced by Digital Signal Processing Synthesizer},''
\newblock in {\em Proc. Interspeech}, 2023.

\bibitem{karaoker}
P. Kakoulidis, N. Ellinas, G. Vamvoukakis, K. Markopoulos, J.~S. Sung, G. Jho,
  P. Tsiakoulis, and A. Chalamandaris,
\newblock ``{Karaoker: Alignment-free singing voice synthesis with speech
  training data},''
\newblock in {\em Proc. Interspeech}, 2022.

\bibitem{wav2vec2}
A. Baevski, Y. Zhou, A. Mohamed, and M. Auli,
\newblock ``{Wav2Vec 2.0: A Framework for Self-Supervised Learning of Speech
  Representations},''
\newblock in {\em Proc. NeurIPS}, 2020, vol.~33.

\bibitem{wavlm}
S. Chen, C. Wang, Z. Chen, Y. Wu, S. Liu, Z. Chen, J. Li, N. Kanda, T.
  Yoshioka, X. Xiao, J. Wu, L. Zhou, S. Ren, Y. Qian, Y. Qian, J. Wu, M. Zeng,
  X. Yu, and F. Wei,
\newblock ``{WavLM: Large-Scale Self-Supervised Pre-Training for Full Stack
  Speech Processing},''
\newblock {\em IEEE Journal of Selected Topics in Signal Processing}, vol. 16,
  no. 6, 2022.

\bibitem{rotary}
J. Su, M. Ahmed, Y. Lu, S. Pan, W. Bo, and Y. Liu,
\newblock ``{RoFormer: Enhanced transformer with Rotary Position Embedding},''
\newblock {\em Neurocomputing}, vol. 568, 2024.

\bibitem{instancenorm}
D. Ulyanov, A. Vedaldi, and V. Lempitsky,
\newblock ``{Instance Normalization: The Missing Ingredient for Fast
  Stylization},''
\newblock {\em arXiv preprint arXiv:1607.08022}, 2017.

\bibitem{mish}
D. Misra,
\newblock ``{Mish: A Self Regularized Non-Monotonic Activation Function},''
\newblock {\em arXiv preprint arXiv:1908.08681}, 2020.

\bibitem{adaspeech}
M. Chen, X. Tan, B. Li, Y. Liu, T. Qin, sheng zhao, and T.-Y. Liu,
\newblock ``{AdaSpeech: Adaptive Text to Speech for Custom Voice},''
\newblock in {\em Proc. ICLR}, 2021.

\bibitem{infovae}
S. Zhao, J. Song, and S. Ermon,
\newblock ``{InfoVAE: Information Maximizing Variational Autoencoders},''
\newblock {\em arXiv preprint arXiv:1706.02262}, 2018.

\bibitem{conformer}
A. Gulati, J. Qin, C.-C. Chiu, N. Parmar, Y. Zhang, J. Yu, W. Han, S. Wang, Z.
  Zhang, Y. Wu, and R. Pang,
\newblock ``{Conformer: Convolution-augmented Transformer for Speech
  Recognition},''
\newblock in {\em Proc. Interspeech}, 2020, pp. 5036--5040.

\bibitem{unet}
O. Ronneberger, P. Fischer, and T. Brox,
\newblock ``U-{N}et: Convolutional networks for biomedical image
  segmentation,''
\newblock in {\em Proc. MICCAI}, 2015.

\bibitem{spectralnorm}
T. Miyato, T. Kataoka, M. Koyama, and Y. Yoshida,
\newblock ``{Spectral Normalization for Generative Adversarial Networks},''
\newblock in {\em Proc. ICLR}, 2018.

\bibitem{gelu}
D. Hendrycks and K. Gimpel,
\newblock ``{Gaussian Error Linear Units (GELUs)},''
\newblock {\em arXiv preprint arXiv:1606.08415}, 2023.

\bibitem{unetd}
T. Kaneko, H. Kameoka, K. Tanaka, and S. Seki,
\newblock ``{Wave-U-Net Discriminator: Fast and Lightweight Discriminator for
  Generative Adversarial Network-Based Speech Synthesis},''
\newblock in {\em Proc. ICASSP}, 2023.

\bibitem{specaugment}
D.~S. Park, W. Chan, Y. Zhang, C.-C. Chiu, B. Zoph, E.~D. Cubuk, and Q.~V. Le,
\newblock ``{SpecAugment: A Simple Data Augmentation Method for Automatic
  Speech Recognition},''
\newblock in {\em Proc. Interspeech}, 2019.

\bibitem{diffusiongan}
Z. Wang, H. Zheng, P. He, W. Chen, and M. Zhou,
\newblock ``{Diffusion-GAN: Training GANs with Diffusion},''
\newblock in {\em Proc. ICLR}, 2023.

\bibitem{realfake}
M. Baas and H. Kamper,
\newblock ``{Disentanglement in a GAN for Unconditional Speech Synthesis},''
\newblock {\em arXiv preprint arXiv:2307.01673}, 2023.

\bibitem{libritts}
H. Zen, V. Dang, R. Clark, Y. Zhang, R.~J. Weiss, Y. Jia, Z. Chen, and Y. Wu,
\newblock ``{LibriTTS: A corpus derived from LibriSpeech for
  Text-to-{S}peech},''
\newblock in {\em Proc. Interspeech}, 2019, pp. 1526--1530.

\bibitem{vctk}
K.~M. Christophe~Veaux, Junichi~Yamagishi,
\newblock ``{CSTR VCTK Corpus: English Multi-speaker Corpus for CSTR Voice
  Cloning Toolkit},'' 2019.

\bibitem{lpcnet}
J.-M. Valin and J. Skoglund,
\newblock ``{LPCNet: Improving Neural Speech Synthesis Through Linear
  Prediction},''
\newblock in {\em Proc. ICASSP}, 2019.

\bibitem{adam}
D.~P. Kingma and J. Ba,
\newblock ``Adam: {A} method for stochastic optimization,''
\newblock in {\em Proc. ICLR}, 2015.

\end{thebibliography}

\end{document}